\documentclass{iau}
\usepackage{amsmath}
\usepackage{amsfonts}
\usepackage{amssymb}
\usepackage{color}
\usepackage{xcolor,graphicx}
\DeclareGraphicsExtensions{.png,.jpg,.pdf}

\title[Estimating the distribution of Galaxy Morphologies on a continuous space] 
{Estimating the distribution of Galaxy Morphologies on a continuous space}

\author[G. Vinci, P. Freeman, J. Newman, L. Wasserman \& C. Genovese]   
{Giuseppe Vinci$^1$, Peter Freeman$^1$, Jeffrey Newman$^2$, \\Larry Wasserman$^1$ \and Christopher Genovese$^1$}

\affiliation{$^1$ Dept. of Statistics, Baker Hall, Carnegie Mellon University, \\ 5000 Forbes Avenue, Pittsburgh, PA 15213, USA \\ email: {\tt gvinci@andrew.cmu.edu} \\[\affilskip]
$^2$Dept. of Physics \& Astronomy, University of Pittsburgh, \\ 310 Allen Hall 3941 O'Hara St., Pittsburgh, PA 15260, USA}

\pubyear{2014}
\volume{306}  
\pagerange{119--126}
\jname{Statistical Challenges in 21$ ^{st} $ Century Cosmology}
\editors{A.C. Editor, B.D. Editor \& C.E. Editor, eds.}

\begin{document}

\maketitle

\begin{abstract}
The incredible variety of galaxy shapes cannot be summarized by human defined discrete classes of shapes without causing a possibly large loss of information.
Dictionary learning and sparse coding allow us to reduce the high dimensional space of shapes into a manageable low dimensional continuous vector space. Statistical inference can be done in the reduced space via probability distribution estimation and manifold estimation.

\end{abstract}

\firstsection 
\section{Introduction}
The evolution of the Universe has led to the formation of complex objects apparently without any regular shape, which our mind would just classify as \textit{irregular}. Thus, the incredible variety of galaxy shapes cannot be summarized by human defined discrete classes of shapes (e.g. ``Hubble sequence'') without causing a possibly large loss of information. Our human concept of shape could limit the complete understanding of the complex structure of the galaxies. Estimating the distribution of galaxy morphologies is one means to test theories of the formation and the evolution of the Universe. We estimate the distribution of morphologies on a \textit{continuous} Euclidean space, such that a particular shape will be viewed as a point in a continuous space. This task must be performed in an \textit{unsupervised} way, i.e. free from any human judgement. Galaxy images are intrinsically high-dimensional data, and we use \textit{dictionary learning} and \textit{sparse coding} [\cite{mai2010}] to reduce the high dimensional space of shapes into a manageable low dimensional one. Essentially, galaxy images will be approximated by sparse linear combinations of basis pictures, which are \textit{learned} from the data. Statistical inference on the reduced space can be performed via probability distribution estimation.
We propose a testing procedure and analyse a dataset of galaxy images\footnote{GOODS-South Early Release Science Field dataset observed in the near-infrared regime by the Wide Field Camera 3 on-board the Hubble Space Telescope [see \cite{wind2011, free2013}].} to show some examples.

\section{Dictionary Learning and Sparse Coding - Radon Transform}
The general idea of dictionary learning and sparse coding is to approximate images by \textit{sparse} linear combinations of a fixed number of \textit{basis images}, which are not predefined, but are \textit{learned} from the data. Let $x_i\in\mathbb{R}^{a\times b}$ be an image, which has $a\times b$ dimensions. For $m<<a\times b$, we want to approximate $x_i$ as:
\begin{equation}\label{approx}
x_i\approx \sum_{j=1}^m\alpha_{ij}B_j
\end{equation}
where $\alpha_i=(\alpha_{i1}, ..., \alpha_{im})\in\mathbb{R}^m$ is a sparse vector of coefficients, and $\{B_j \}_{j=1}^m $ is a collection of basis images $B_j\in \mathbb{R}^{a\times b} $.
Notice that the basis images will not be imposed to be orthogonal
such that the dictionary can easily adapt to the structure of the data [\cite{mai2010}]. Moreover, learning the bases from the data was shown to perform better in signal reconstruction with respect to using predefined bases [\cite{elad2006}].

\subsection{Optimization problem}\label{optim}
From a dataset of galaxy images $ \{x_i\}_{i=1}^n $, we can estimate the dictionary $D =\{B_j\}_{j=1}^m $ and the vectors of coefficients $A=\{\alpha_i\}_{i=1}^n $ by solving the following optimization problem:
\begin{equation}\label{optprob}
\left\{\begin{array}{lll}
\underset{\{{\alpha_i}\}_{i=1}^n,  \{{B_j}\}_{j=1}^m }{\min}\sum\limits_{i=1}^n\left[\dfrac{1}{2} \left \Vert x_i- \sum\limits_{j=1}^m{\alpha_{ij}B_j}\right\Vert_2^2+\underbrace{\lambda\Vert{\alpha_i}\Vert_1}_{\textsc{Sparsity}}\right]\\
\textbf{s.t.}\hspace*{4mm} \Vert{B_j}\Vert_2^2 \leq 1, \forall j=1, ...,m
\end{array}\right.
\end{equation}
where $\lambda\geq 0$ is a sparsity parameter and $ \Vert *\Vert_2^2 $ is the Frobenius norm [\cite{mai2010}; \textsf{R} package ``\textsf{spams}'']. We suggest to choose $m$ and $\lambda $ via \textit{cross validation}. See \cite{mai2010} for other configurations of problem (\ref{optprob}).

\subsection{Standardization of the images. Radon transform}\label{stand}
\begin{figure}[ht!]
  \centering
\includegraphics[width=1\textwidth]{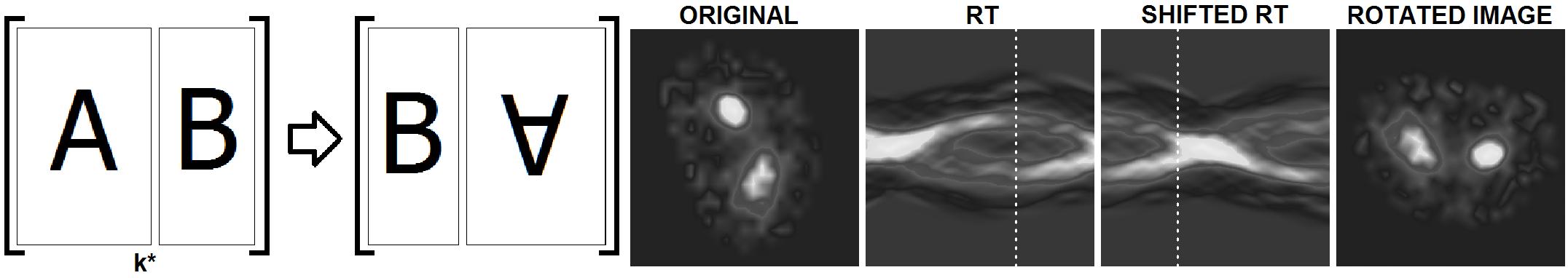}     
 \caption{\small Left: vectors $A$ are moved after vectors $B$ with values moved up and down. Right: starting from an original image, we compute its Radon transform on a discrete grid, then by shifting the vectors of this matrix according to the orientation $\theta^* $, we can obtain a standardized rotated version of the image as the IRT of the shifted RT.}
  \label{radrot}
\end{figure}

\begin{figure}[ht!]
  \centering
\includegraphics[width=0.59\textwidth]{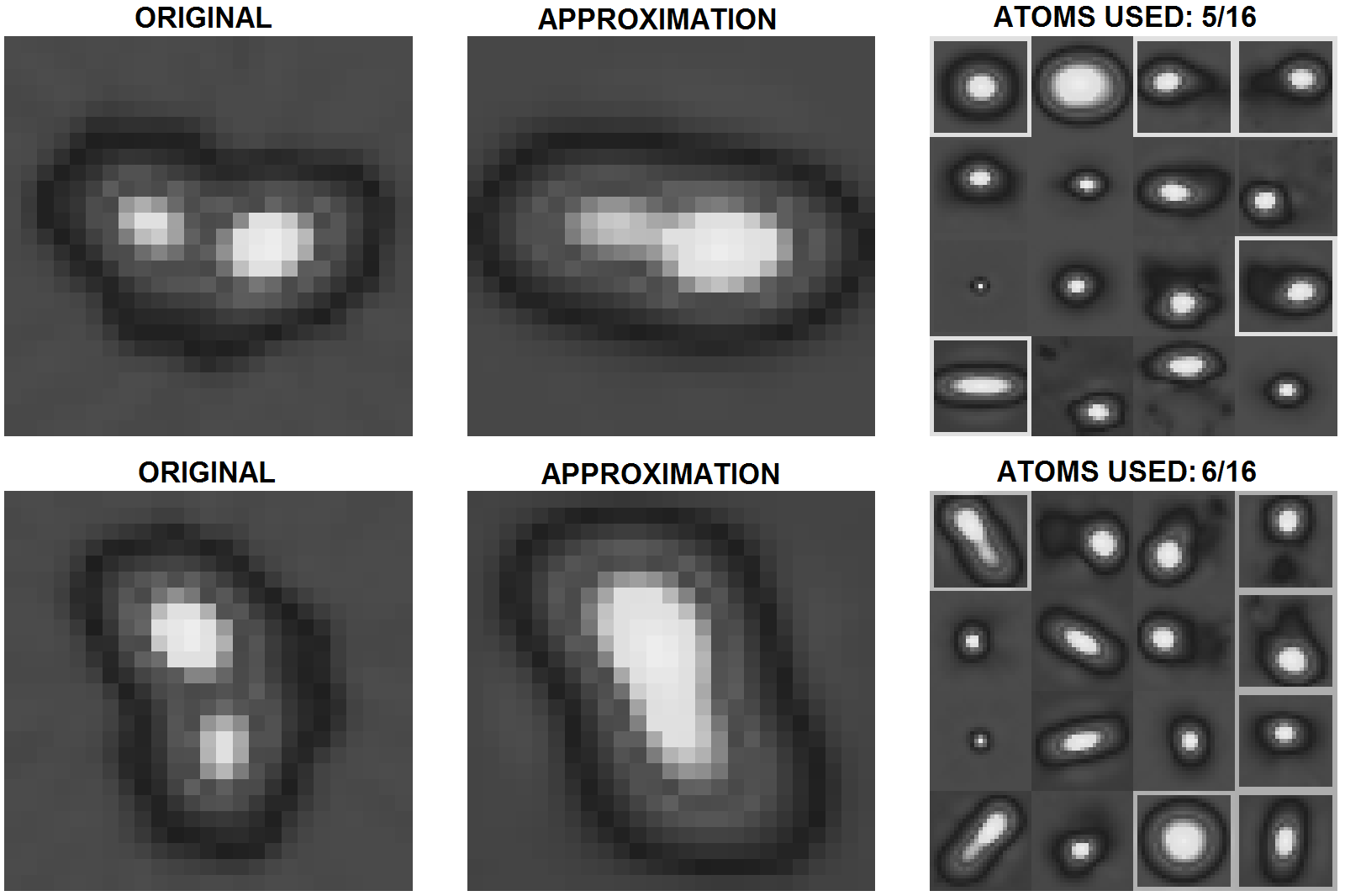}          \includegraphics[width=0.4\textwidth]{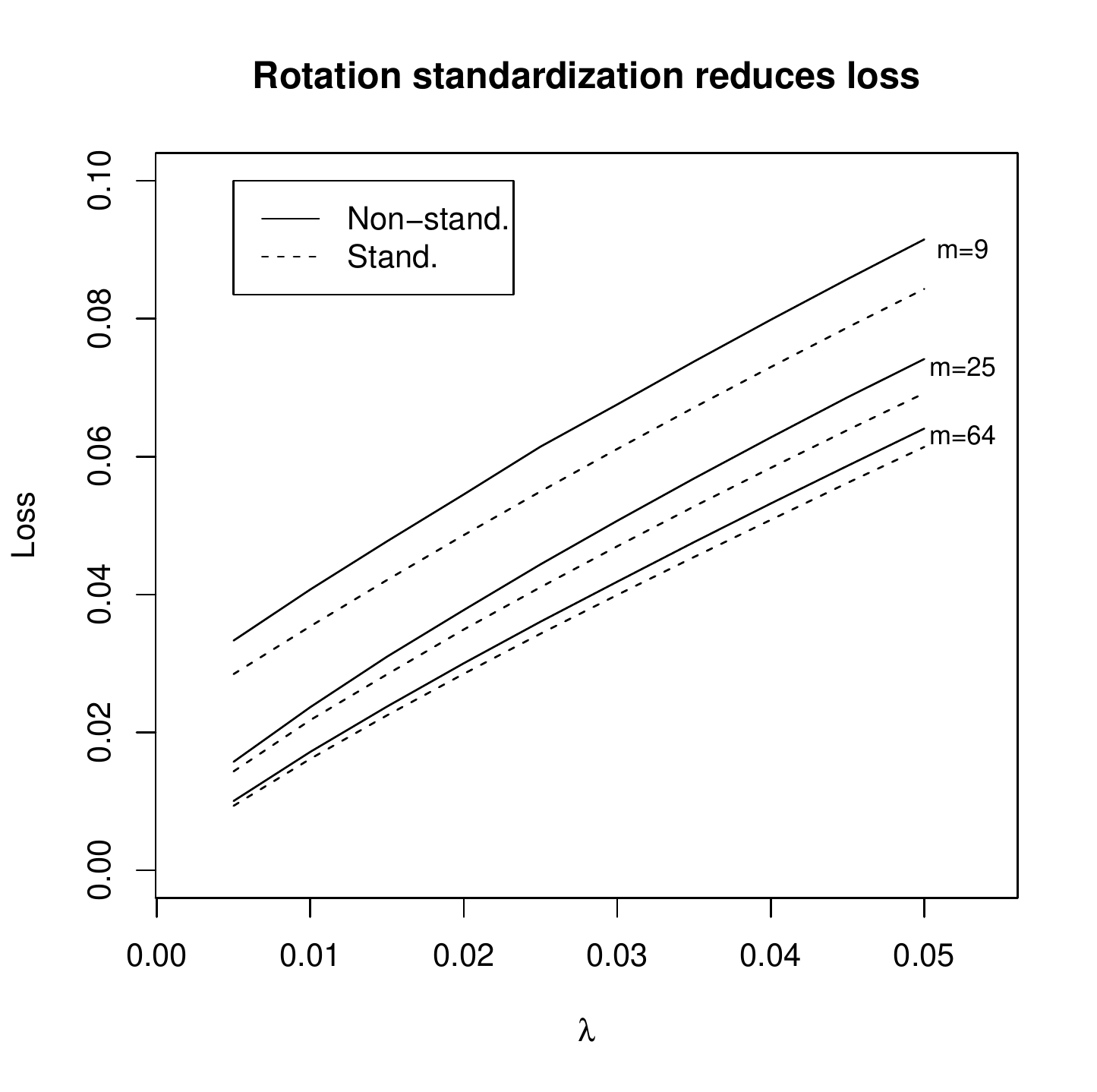}      \caption{\small Rotation standardization improves the fit. Left: an image approximated using a dictionary learned with rotation standardization (top) and not (bottom). Spurious dimensionality negatively affects the dictionary at the bottom, while rotation standardization may lead to more refined approximations. Right: for different numbers of atoms ($m=9, 25, 64$), the minimum loss (\ref{optprob}) is smaller when using standardized images. Images are from the GOODS-S dataset, H-band.}
  \label{sqor}
\end{figure}

Before solving problem (\ref{optprob}), images must be standardized to eliminate any spurious dimensionality and improve the quality of the approximations (\ref{approx}). We are talking about: \textit{centring}, \textit{resizing} and \textit{rotation orientation}. While the first one can be easy to perform, the two others are not. Images can be rotated and resized by using Radon Transform (RT) and Inverse RT (IRT). The RT of a function $f$ is $\mathcal{R}_f(t,\theta)= \int_{-\infty}^\infty f(t\cos\theta-u\sin\theta, t\sin\theta+u\cos\theta)du$, where $(t,\theta)\in \mathbb{R}^2 $. An image can be viewed as the discrete evaluation of a function. The orientation of the texture of an image can be estimated by $\theta^*=\arg\min_\theta \frac{\partial^2\sigma^2_\theta}{\partial \theta^2} $, where $\sigma_{\theta}^2 $ is the variance of $\mathcal{R}_f(t,\theta)$ at angle $\theta$ [\cite{jaf2005, arod2012}; \textsf{R} package ``\textsf{PET}'']. Rotating images by angle $-\theta^* $ essentially makes all the pictures \textit{horizontally oriented}. To rotate an image we need to: 1) evaluate its RT on a discrete grid, say $\hat{R}_{M\times (\omega 180+1)} =\{\mathcal{R}_f(t,\theta) \}$ with $t\in\{t_1,...,t_M\}$, $\theta\in \{\frac{j}{\omega 180}\pi \}_{j=0}^{\omega 180}$, and $\omega\in\mathbb{N}^+$; 2) find $\theta^*$ and move the first $k^*=\theta^*\frac{\omega 180}{\pi} $ columns of $\hat{R}$ as described in Figure \ref{radrot} to get $\tilde{R}$ (``rotation'' in the Radon domain); 3) computing the IRT of $\tilde{R}$ on a grid of desired resolution (``resizing''). In Figure \ref{sqor} we show some effects of images standardization.

\section{Statistical inference on the reduced space}
In this section we propose a method to estimate the distribution of galaxy morphologies on a low-dimensional space, and we use the GOODS-S dataset to perform a simulation. 

\subsection{Probability distribution of galaxy morphologies.}
For a dataset of $n$ images $ \{x_i \}_{i=1}^n $, where $ x_i $ is a matrix of nonnegative light intensity:
\begin{enumerate}
\item Standardize all the images as described in paragraph \ref{stand};
\item Obtain the dictionary $D$ and the vectors $A=\{\alpha_i\}_{i=1}^n$ according to paragraph \ref{optim};
\item Estimate the joint distribution of vector $\alpha_i \in \mathbb{R}^m$. Call it $ \hat{P}_\alpha$. 
\end{enumerate}
Given the fitted dictionary $D$, estimate $ \hat{P}_\alpha$ can be viewed as an approximation of the distribution of galaxy morphologies.

\subsection{Comparing populations of shapes}\label{simuls}
In this section we propose a method to compare the distributions of two collections of images. Let $ X, Y $ be two collections of images. Suppose we want to test hypothesis $X\stackrel{\mathcal{D}}{=}Y $, i.e. a distribution test. We propose the following method:
\begin{enumerate}
\item Pool $X$ and $Y$ into a unique dataset $Z=[X,Y]$
\item From $Z$, fit dictionary $D$ and vectors of coefficients $\{\alpha_{Z,k}\}=[\{\alpha_{X,i}\},\{\alpha_{Y,j}\}]$. 
\item Implement a distribution test $ \alpha_X\stackrel{\mathcal{D}}{=}\alpha_Y $.
\end{enumerate}
For step $(c)$, we suggest to use the nonparametric test based on the Maximum Mean Discrepancy (MMD) statistic (\cite{gre2012}; \textsf{R} package ``\textsf{kernlab}''). We can call this testing procedure ``DSM test'' (Dictionary Learning - Sparse Coding - MMD).

\subsubsection{Simulation}
\begin{figure}[ht!]
  \centering
 \includegraphics[width=1\textwidth]{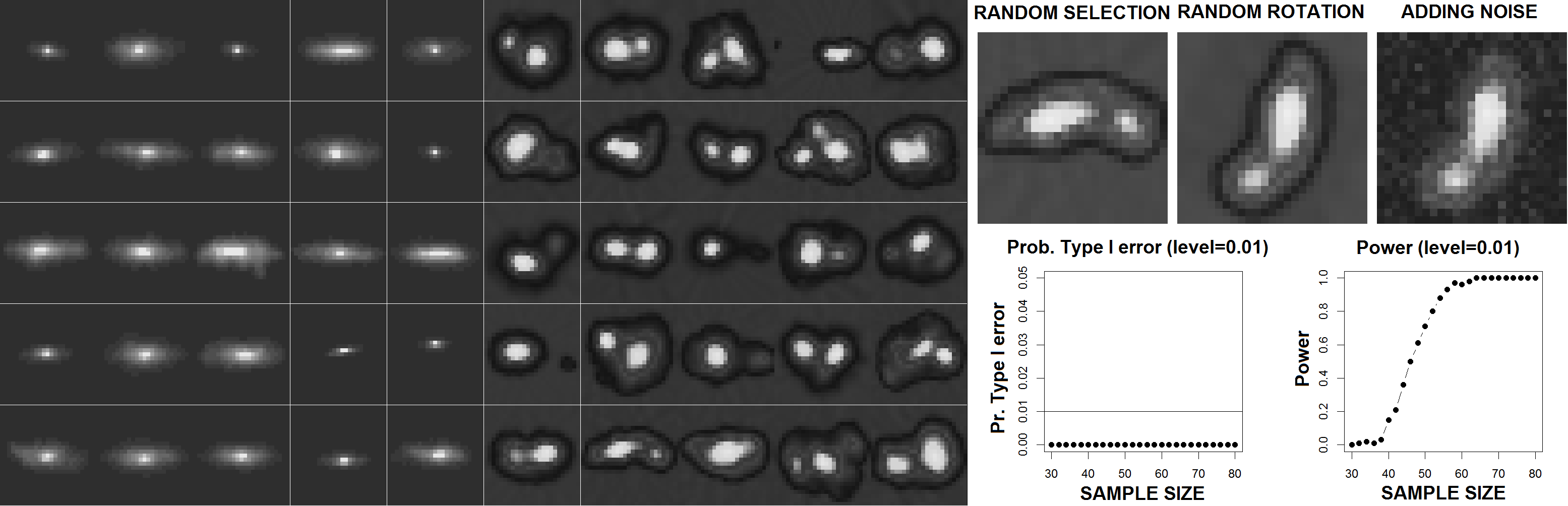}
    \caption{\small Left: selected non-mergers ($X_1$) and mergers ($X_2$) from the GOODS-S dataset, H-band. Top right: procedure to simulate an image from $X_i$. An image is randomly selected from the subset, randomly rotated and heteroscedastic Gaussian noise is added to each pixel. Bottom right: the DSM test helps to distinguish different shapes. The probability of Type I error of the DSM test is always smaller than the level of the test; the power of the test is increasing in the sample size. The shape of the power function depends on the original sets $X_1, X_2$.}
  \label{simplots}
\end{figure}

We selected two subsets of images of the GOODS-S dataset in the H-band (see Figure \ref{simplots}): $X_1$ with 25 images of non-mergers, and $X_2$ with 25 images of mergers. To generate $n$ images of non-mergers and $n$ images of non-mergers we: 1) randomly sample with replacement $n$ images from $X_1$ and $n$ images from $X_2$, respectively; 2) randomly rotate them by angles $\theta\sim$Unif$(0, 2\pi)$, i.i.d.; 3) add heteroscedastic noise: $\epsilon_{jk}\stackrel{\text{indep}}{\sim} N(0,\beta^2\times I_{jk} ) $, where $I_{jk}\geq 0$ is the light intensity at position $jk$ in a matrix. We repeat comparisons (via DSM test) of samples of the same kind (Mer Vs Mer, NMer Vs NMer) and different one (Mer Vs NMer) to estimate the probability of Type I error and the power of the test as functions of the sample size (see Figure \ref{simplots}).
We chose $m=4$ and $\lambda=0.05$ via 10-CV.

\section{Conclusions and future work}
An unsupervised analysis based on dictionary learning and sparse coding allows us to approximate the distribution of galaxy morphologies by a multivariate distribution defined on a subset of $\mathbb{R}^m $, where dimension $m$ is much smaller than the dimension of a galaxy image. Hypothesis testing on the reduced space can help to distinguish the distributions of two sets of images. Current and future work is: using dictionary learning and sparse coding to put constraints on the parameters of cosmological models; comparing the distribution of galaxy shapes at different redshift ranges; manifold estimation: some clusters may correspond to some human defined shapes (e.g. spiral, elliptical) and filaments [see \cite{chen2013}] may describe the transition from a shape to another one; analysing images of other astronomical objects and 3D images.

\end{document}